\documentclass[a4paper,prl,reprint]{revtex4-1}
\usepackage{graphicx}
\usepackage{amsmath}
\usepackage{fancyhdr}
\usepackage{cancel}
\usepackage{color}
\setcitestyle{round}
\begin{document}
\title{Stationary properties of maximum entropy random walks}
\author{Purushottam D. Dixit}
\affiliation{Department of Systems Biology, Columbia University\\ New York, NY}
\thanks{Email: pd2447@c2b2.columbia.edu}
%
%\author{Ken A. Dill}
%\affiliation{Laufer Center for Physical and Quantitative Biology,\\Department of Chemistry,\\ and Department of Physics and Astronomy,\\Stony Brook University,\\ Stony Brook, NY}

\begin{abstract}
Maximum entropy (maxEnt) inference of state probabilities using state-dependent constraints is popular in the study of complex systems. In stochastic dynamical systems,  the effect of state space topology and path-dependent constraints on the inferred state probabilities is unknown. To that end, we derive the  transition probabilities and the stationary distribution of a maximum {\it path} entropy Markov process subject to state- and path-dependent constraints. The stationary distribution reflects a competition between path multiplicity and imposed constraints and is significantly different from the Boltzmann distribution. We illustrate our results with a particle diffusing on an energy landscape.  Connections with the path integral approach to diffusion are discussed.  
\end{abstract}
\maketitle

Owing to our increasing ability to collect large amounts of data in complex systems and our inability to construct {\it generative} models to explain that data, {\it descriptive} approaches have become popular. One such framework is the principle of maximum entropy (maxEnt)~\citep{it_theory1,gibbs,shore,Press2012}.  Intuitively, maxEnt picks the `least informative' distribution over states while requiring it to reproduce certain aspects of the data. The result is the Boltzmann distribution in constrained quantities. maxEnt has been employed to study a variety of problems, for example,  neuronal firing patterns~\citep{schneidman2006weak}, bird flocks~\citep{bialek2012statistical,cavagna2014dynamical}, ecological species  distribution~\citep{phillips2006maximum}, gene expression noise~\citep{dixit2013quantifying}, sequence variability in proteins~\citep{mora2010maximum,shekhar2013spin},  and behavior~\citep{peterson2013maximum}.

In many cases~\citep{phillips2006maximum,schneidman2006weak,bialek2012statistical,dixit2013quantifying,cavagna2014dynamical}, but not always~\citep{mora2010maximum,shekhar2013spin,peterson2013maximum}, the experimental data is a realization of a stochastic process. In such cases, one may wish impose path-dependent current like constraints in addition to state-dependent constraints. Moreover, the dynamical radius of any state --- the states reachable in a single transition --- is usually finite, which defines the state space topology. How these factors affect inferred state probabilities is unknown. 

We solve this problem for Markovian dynamics in discrete state and time. In order to incorporate dynamical information, we maximize a {\it path} entropy. We derive transition probabilities and the stationary distribution of the maximum path entropy Markov process subject to state- and path-dependent constraints.  The stationary distribution is the product of the left and the right Perron-Frobenius eigenvectors of a matrix and depends non-trivially on the topology and imposed constraints.  We illustrate our results with a random walk diffusing on a two dimensional energy landscape. 

We begin with an observation. Discrete state stochastic systems can be modeled by a random walk in higher dimensions. For example, the time evolution of an Ising model with $N$ spins is a random walk in $2^N$ dimensions. If at most one spin flip per transition is allowed, for example the popular Glauber dynamics~\citep{glauber1963time}, every state is connected to only $N$ out of the $2^N$ states.  To that end, we consider an irreducible and aperiodic discrete time Markovian random walk on a directed graph $G$ with nodes $V$ and edges $E$. We denote the {\it unique} stationary distribution over the states by $\{ p_a \}$.   We assume that transition probabilities $k_{ab} \neq 0$ only when $(a,b) \in E $. 

We seek the maximum entropy stationary distribution subject to state- and path-dependent constraints. The appropriate ensemble to impose these constraints is the ensemble $\{ \Gamma \}$ of stationary state trajectories $\Gamma \equiv \cdots \rightarrow a \rightarrow b \rightarrow \cdots$ of fixed but unspecified duration $T$. We only consider trajectories that are permissible by the state space topology. The entropy of the ensemble, normalized by $T$, is given by~\citep{dixit2014inferring,dixit2015inferring,filyukov1967method,cover2012elements}
\begin{eqnarray}
\mathcal S &=& -\frac{1}{T}\log P(\Gamma) \log P(\Gamma) = -\sum_{a,b} p_a k_{ab} \log k_{ab}\label{eq:pathentropy}
\end{eqnarray}
In Eq.~\ref{eq:pathentropy} and from here onwards, unless speciefied otherwise, all summations involving quantities with two indices are restricted on the edges of the graph. 

$\{ p_a \}$ and $\{ k_{ab} \}$ are not independent of each other. In fact, they are constrained as follows
\begin{eqnarray}
\sum_b p_a k_{ab} &=& p_a,~\sum_a p_a k_{ab}= p_b, ~\sum_{a,b} p_a k_{ab} = 1. \label{eq:stationary}
\end{eqnarray}
If the dynamics is reversible, the walk also satisfies detailed balance conditions,
\begin{eqnarray}
p_a k_{ab} = p_b k_{ba}.\label{eq:db}
\end{eqnarray}

Let us introduce constraints of path ensemble averages of state- and path-dependent quantities $r^i_{ab}$.  State-dependent quantities $r^{i}_{ab}$ such as energy and particle number depend only on the initial state $a$ or the final state $b$. Path-dependent quantities $r^{i}_{ab}$ such as energy or particle currents depend on both states.  The path ensemble averages are given by~\citep{dixit2014inferring,filyukov1967method,dixit2015inferring}
\begin{eqnarray}
       \langle r^i \rangle = \sum_{a,b} p_a k_{ab} r^{i}_{ab}. \label{eq:const}
\end{eqnarray}

We maximize the path entropy $\mathcal S$ in Eq.~\ref{eq:pathentropy} with respect to unknown stationary distribution $p_a$ and transition probabilities $k_{ab}$ while imposing constraints  in Eqs.~\ref{eq:stationary} and Eq.~\ref{eq:const}.  Using Lagrange multipliers, we write the unconstrained Lagrange function, sometimes called the Caliber~\citep{Stock2008,Press2012},
\begin{eqnarray}
\mathcal C &=&\mathcal S + \sum_a m_a \left (\sum_b p_a k_{ab} - p_a \right ) +   \sum_b n_b \left (\sum_a p_a k_{ab} - p_b \right ) \nonumber \\ &+&    \delta \left ( \sum p_a k_{ab} - 1 \right )  - \sum_i \gamma_i \left (\sum_{a,b}\left (p_a k_{ab} r^i_{ab}- \langle r^i \rangle \right ) \right ). 
\end{eqnarray}

Maximizing the Caliber with respect to  $p_a$ and $k_{ab}$, we find that the transition probabilities $k_{ab}$ are given by (see appendix for details)
\begin{eqnarray}
k_{ab} = \frac{1}{\eta}{\frac{\phi_b}{\phi_a}} {\bf W}_{ab} \label{eq:main1}
\end{eqnarray}
where the elements of the {\it constraint matrix} $\bf W$ are given by
\begin{eqnarray}
 {\bf W}_{ab} = \exp \left (-\sum_i \gamma_i r^i_{ab} \right )
\end{eqnarray} when $(a,b) \in E$ and zero otherwise.   $\bar \phi$  is the normalized eigenvector of ${\bf W}$ corresponding to its maximum eigenvalue $\eta$. The Perron-Frobenius theorem guarantees that $\bar \phi$ is strictly positive and $\eta$ is unique and positive. A simple case of Eq.~\ref{eq:main1} for a freely diffusing random walk was studied by Burda et al.~\citep{burda2009localization} where ${\bf W}$ is equal to the adjacency matrix of the graph $G$.

The stationary distribution $\{p_a\}$ can be determined by solving the linear system of equations
\begin{eqnarray}
\sum_a p_a k_{ab} = p_b \Rightarrow \sum_a \frac{p_a}{\phi_a}{\bf W}_{ab} = \eta \frac{p_b}{\phi_b}.
\end{eqnarray}
Thus, if $\bar \psi$ is the left Perron-Frobenius eigenvector and $\bar \phi$ is the right Perron-Frobenius eigenvector of ${\bf W}$ with the same eigenvalue $\eta$, the stationary distribution is given by the product
\begin{eqnarray}
p_a = \psi_a \phi_a. \label{eq:statprob}
\end{eqnarray} The Perron-Frobenius eigenvectors and thus the stationary distribution depend on the topology and the imposed constraints in a non-trivial fashion.  In other words, the Boltzmann distribution, obtained by maximizing the entropy over state-distributions, is no longer guanranteed when dynamical information is introduced.

\begin{figure}[h]
	\includegraphics[scale=0.4]{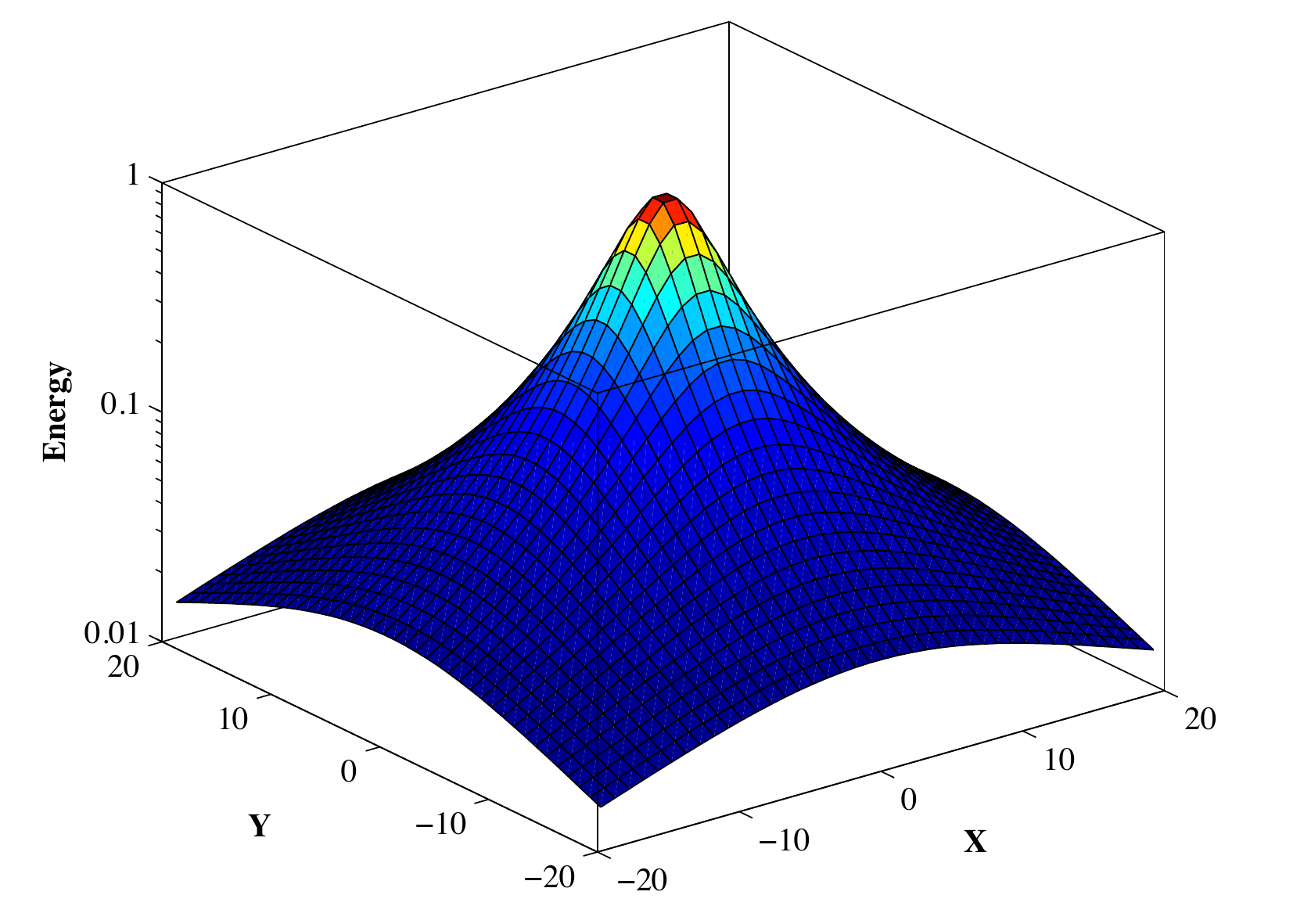}
	\caption{Energy landscape on a $N\times N$ square lattice with $N=40$. Energy is  heighest at the center of the lattice and decreases as the reciprocal of the squared distance from the center (see Eq.~\ref{eq:enls}). We have chosen $A=11$ and $B = 10$. \label{fg:energy_landscape}}
\end{figure}

Is the inferred Markov process reversible? Let us calculate its entropy production rate $\dot s$~\citep{schnakenberg1976network},
\begin{eqnarray}
\dot s = \sum_{a,b} p_a k_{ab} \log \frac{k_{ab}}{k_{ba}} = -\sum_i \gamma_i \langle  r^i_{ab} - r^i_{ba} \rangle. \label{eq:epr}
\end{eqnarray}
In Eq.~\ref{eq:epr}, only the antisymmetric part of constraints $r^i_{ab}$ contributes to entropy production. If all constraints are symmetric, the entropy production is zero and the Markov process is reversible. In fact, if microscopic reversibility (Eq.~\ref{eq:db}) is explicitly imposed, the inference problem is equivalent to constraining  symmetrized quantities $r^{i\dag}_{ab} = \frac{1}{2} \left ( r^i_{ab} + r^i_{ba} \right )$ (see appendix for details). In this case, the constraint matrix ${\bf W}$ is symmetric and the left and the right Perron-Frobenius eigenvectors coincide. The stationary distribution is simply the square of this eigenvector.

Finally, we write down the probability of an arbitrary path $\Gamma = a_1\rightarrow a_2 \rightarrow a_3 \rightarrow \cdots \rightarrow a_n$ of total duration $n$. If the initial state $a_1$ is chosen from a distribution $p_0(a_1)$, we have
\begin{eqnarray}
p(\Gamma) &=& p_0(a_1) \cdot k_{a_1a_2} \cdot k_{a_2a_3} \cdots k_{a_{n-1}a_n} \\
&=& \frac{p_0(a_1)}{\phi_{a_1}} \frac{1}{ \eta^{n-1}} e^{-\mathcal A(\Gamma)} \label{eq:action}
\end{eqnarray}
where $\mathcal A(\Gamma)$ is the `action' associated with the path $\Gamma$ and is given by
\begin{eqnarray}
\mathcal A(\Gamma) = \sum_i \gamma_i \sum_{t=1}^{n-1} r^i_{a_ta_{t+1}}.
\end{eqnarray}

Our construction of the maximum path entropy Markov process and its stationary distribution is complete. While it gives us a recipe to calculate the stationary distribution, Eq.~\ref{eq:statprob} does not allow us an intuitive understanding of how it  depends on topology and constraints. Below, we will illustrate three important features that are uniqe to path entropy maximization, path entropy/enthalpy compensation, state space topology, and currents. 

In an illustrative example, we consider a particle diffusing on a $N\times N$ square lattice. In a single transition, the particle jumps to one its nearest neighbors. We define the energy at every point $a = (x,y)$ as
\begin{eqnarray}
\epsilon_{a} = \frac{A}{x^2+y^2+B}. \label{eq:enls}
\end{eqnarray}
$A$ and $B$ are positive constants. Below, we fix $A=11$ and $B=10$. The energy function is symmetric in $x$ and $y$, has a peak in the middle of the lattice, and takes its lowest values in the four corners (see Fig.~\ref{fg:energy_landscape}).

\begin{figure}
	\includegraphics[scale=0.3]{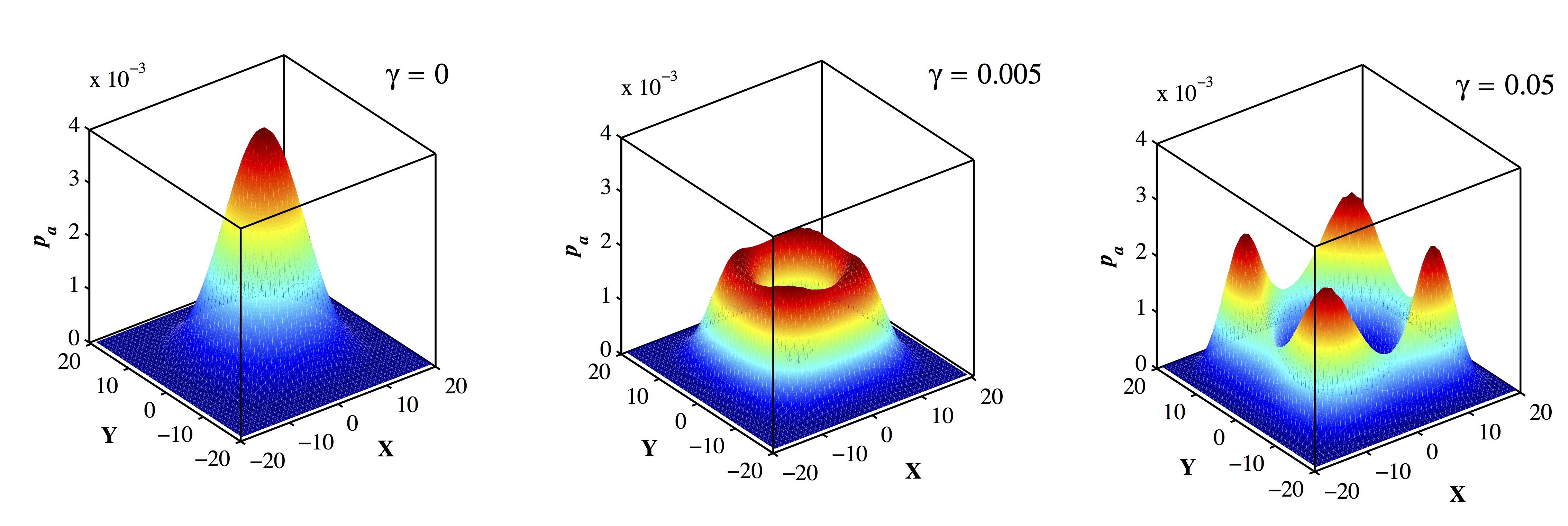}
	\caption{Stationary probabilities $p_a$ in a finite square lattice when average energy constraints are imposed. The particle  localizes in the center of the lattice in the absence of constraints ($\gamma = 0$, left panel). When average energy constraints are used, the particle finds a balance between multiplicity of paths and energetics of the states (center and right panels).  \label{fg:probabilities}}
\end{figure}

First, let us assume that the square lattice is {\it aperiodic}. Corner points, edges, and interior points  have 2, 3, and 4 nearest neighbors respectively. Let us obtain the stationary distribution with constraints of average energy and detailed balance. We first construct the {\it symmetric} constraint matrix 
\begin{eqnarray}
{\bf W}_{ab} = \exp \left [ -\gamma \left ( \frac{\epsilon_a + \epsilon_b}{2} \right ) \right ] \label{eq:results1}
\end{eqnarray}
when $a$ and $b$ are nearest neighbors on the lattice and zero otherwise. $\gamma$ is the Lagrange multiplier associated with the average energy constraints. We then find $\bar \phi$,  its right Perron-Frobenius eigenvector. The stationary distribution is $p_a \propto \phi_a^2$.

In Fig.~\ref{fg:probabilities} we show the stationary distribution for $\gamma = 0, 0.005,$ and $0.05$.  $\gamma = 0$ is denotes  absence of energy constraint.  In this case, the particle localizes near the center of the lattice, a striking departure from the microcanonical maxEnt distribution which predicts equal probabilities for all states. The entropic localization results from the higher multiplicity of paths in the central region compared to the boundaries~\citep{burda2009localization}. When average energy constraints are imposed ($\gamma > 0$), the particle balances the entropic multiplicity of paths with energetic unfavorability of states. This balance  is remniscent of entropy/enthalpy compensation~\citep{lumry1970enthalpy} well known in chemistry. At $\gamma = 0.05$, the particle spontaneously localizes in one of the four corners. Instead of choosing low energy regions near the vertical and horizontal boundaries, the particle chooses regions near the diagonals because of their higher path multiplicity.

\begin{figure}
	\includegraphics[scale=0.4]{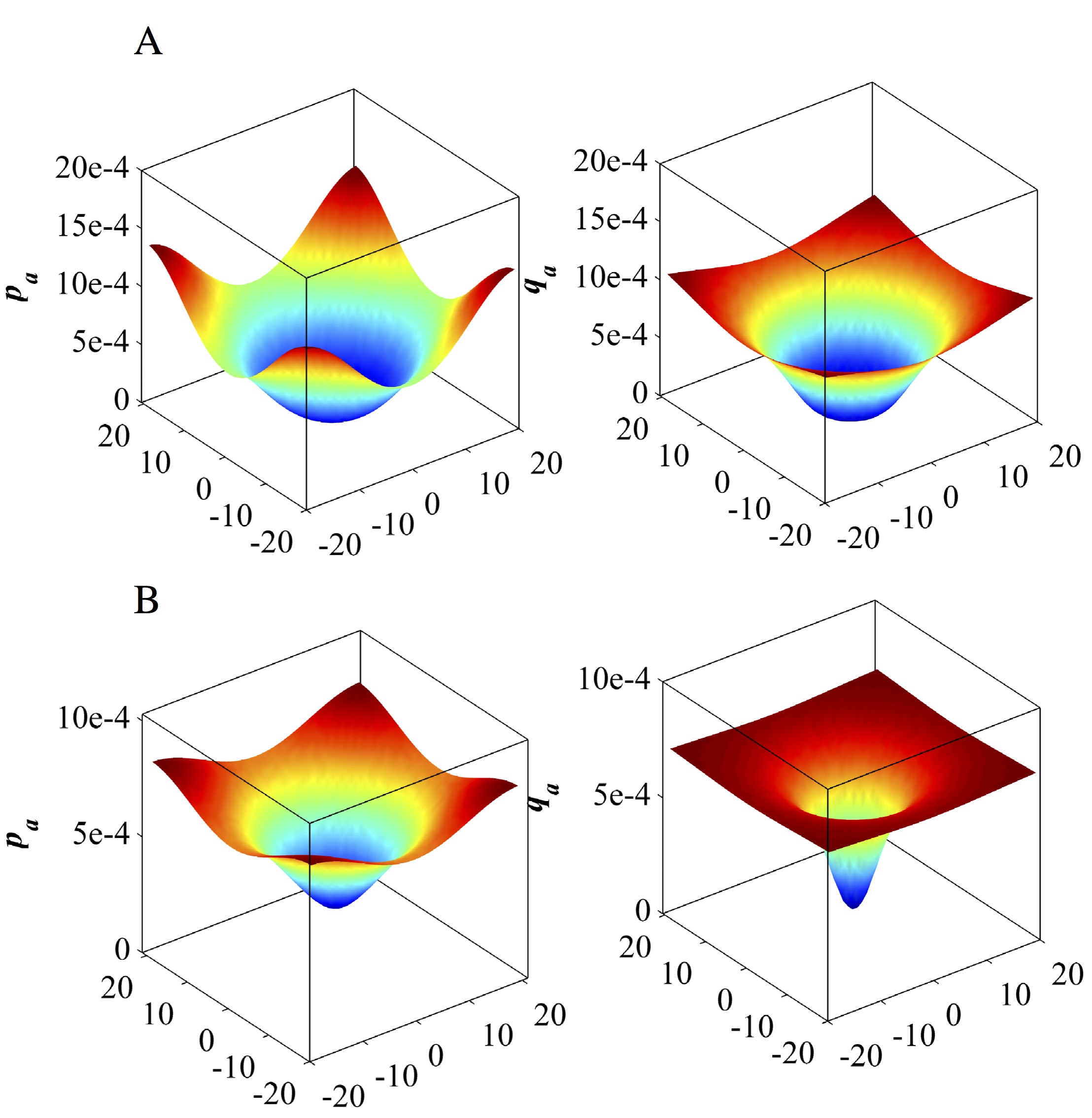}
	\caption{The maximum path entropy stationary distribution $p_a$ and  the Boltzmann distribution $q_a$ when average energy constraints are imposed.  The particle is allowed to jump to the nearest neighbor (A, top) and up to the third nearest neighbor (B, bottom).\label{fg:finite_ratio}}
\end{figure}

Thus, asymmetry in state space topology has a huge impact on the stationary distribution. Are state-based maxEnt and maximum path entropy distributions equal when all states are topologically equivalent?  We give the answer in the negative. Consider a {\it periodic} $N\times N$ square lattice. The only topological restriction is that in a single time step, the particle is allowed to jump to only a finite number of states.

\begin{figure*}
	\includegraphics[scale=0.3]{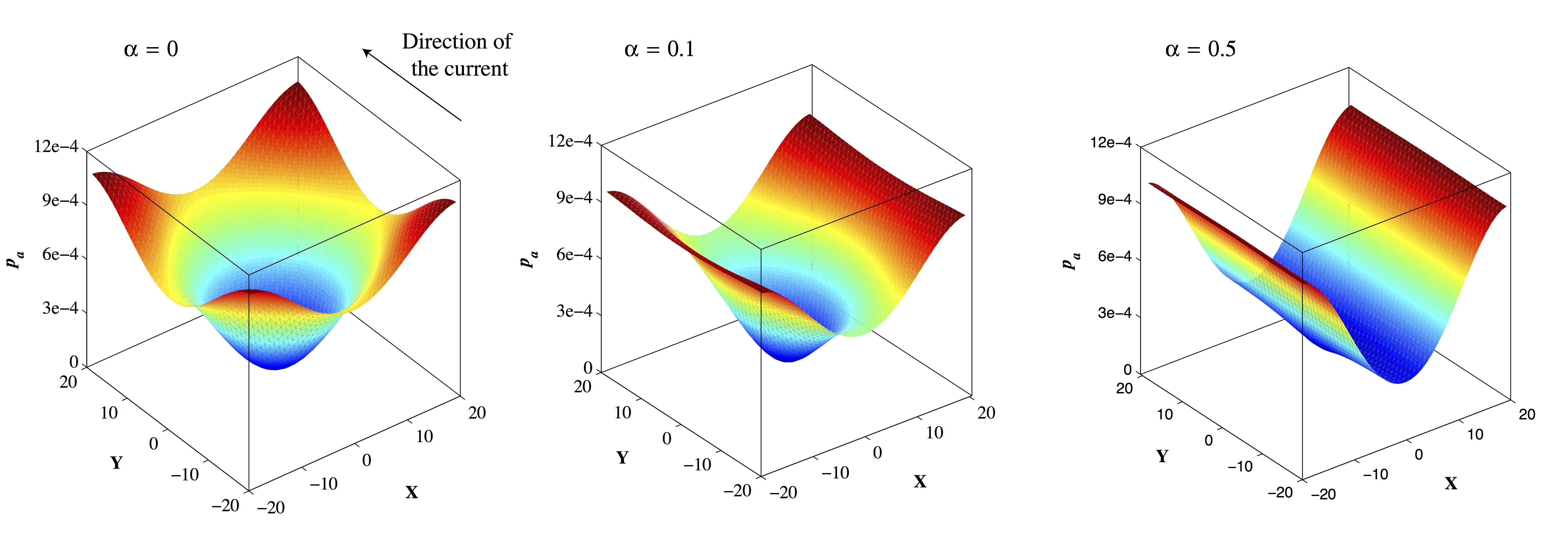}
	\caption{The change in the maximum path entropy stationary distribution in the presence of non-equilibrium current.  Net currents across the boundaries of a system will allow regions of high energy to be frequently visited and vice versa for regions of low energy. As $\alpha$ increases (from left to right), the stationary probability of states near $Y=0$ and $X = \pm 20$ increases and the probability of states near $Y=\pm 20$ and $X=0$ decreases. \label{fg:currents}}
\end{figure*}

In Fig.~\ref{fg:finite_ratio} we plot the stationary distribution $p_a$ (Eq.~\ref{eq:statprob}), the Boltzmann distribution $q_a \propto e^{-\beta \epsilon_a}$, and their ratio after constraining the mean energy.  $p_a$ is calculated  as above with a slight modification that the underlying graph of connectivity represents a periodic lattice.  $\gamma$ (see Eq.~\ref{eq:results1}) is fixed at 0.025.  Inverse temperature $\beta$ is adjusted to match the numerical value of the mean energy, which allows a direct comparison.  We study two different state space topologies. On the top (A), we allow the particle to jump to any one of its nearest neighbors in a single transition. On the bottom (B), we allow the particle to jump up to three Hamming distance away. In both cases, $p_a$ is significantly different than $q_a$ especially in the region of high energy. How do we understand this difference? On the one hand, the maxEnt distribution $q_a$ depends solely on the state energy $\epsilon_a$. On the other hand, Eq.~\ref{eq:action} shows that the paths that visit states of both high and low energy have a non-negligible probability thereby increasing the stationary probability $p_a$ of high energy states compared to $q_a$. As the dynamical reach of the particle is increased from first nearest neighbor to third nearest neighbor, the difference between the maxEnt distribution and the maximum path entropy distribution gets smaller; mean of the absolute log ratio of the probabilities decreases from $\sim0.75$ to $\sim 0.5$ (0 for identical distributions). Indeed, if the particle can jump from any state to any other state in a single transition, the maxEnt and the maximum path entropy predictions are trivially identical to each other~\citep{filyukov1967method}.

In addition to state-dependent quantities like energy, one may wish to constrain path-dependent quantities, like currents. How do path-dependent constraints change the stationary distribution? Let us consider the periodic $N\times N$ square lattice as above. We constrain the average energy and a current along the positive Y axis (see Fig.~\ref{fg:energy_landscape} and Fig.~\ref{fg:currents}).  To obtain the stationary distribution, we first identify the {\it asymmetric} constraint matrix
\begin{eqnarray}
{\bf W}_{ab} = \exp \left [ -\gamma \left (\frac{\epsilon_a + \epsilon_b}{2} \right ) - \alpha J_{ab}\right ].
\end{eqnarray}
As above, $\gamma$ is the Lagrange multiplier associated with energy and $\alpha$ is associated with current. The current in the positive Y direction between states $a = (x,y)$ and $b = (z,w)$ is defined as $J_{ab} = \pm 1$ if $w = y  \pm 1$ with appropriate corrections at $y, w = 1, N$. $J_{ab}$ is zero for sideways movement.  Note that $J_{ab}$ is antisymmetric and contributes to entropy production. We find the left and the right Perron-Frobenius eigenvectors $\bar \psi$ and $\bar \phi$ of ${\bf W}$. The stationary distribution is the product of these two vectors, $p_a = \psi_a \phi_a$.

Fig.~\ref{fg:currents} shows the stationary distribution at $\alpha = 0, 0.1,$ and $0.5$ and $\gamma$ held fixed at $\gamma = 0.025$. At $\alpha=0$, there are no net currents and the stationary distribution is governed entirely by the energy constraints. When we increase $\alpha$ to 0.1 (center) and 0.5 (right), we see that net currents modulate the stationary distribution, a fact well known in statistical physics~\citep{Kon1998}. This effect can be understood by looking at path probabilities. From Eq.~\ref{eq:action}, we know that paths that traverse through high energy regions have a low probability. But, this may be alleviated if they simultaneously carry a net favorable current.  This leads to a higher probability for energetically unfavorable states that are represented frequently in current carrying paths.

In summary, Fig~\ref{fg:probabilities}, Fig.~\ref{fg:finite_ratio}, and Fig.~\ref{fg:currents} show that asymmetry in state space topology, finite dynamical reach of states, and path-dependent constraints all can alter the inferred stationary distribution in a non-trivial fashion.  These effects will likely be magnified in higher dimensions and are relevant in many  discrete state systems where state-based maxEnt has previously been employed~\citep{schneidman2006weak,phillips2006maximum,bialek2012statistical,dixit2013quantifying,cavagna2014dynamical}.  It will be interesting to see whether these additional features lead to better predictive models.

We discussed how dynamical information affects the estimate of the inferred state probabilities. But, we also have access to the path probabilities (see Eq.~\ref{eq:action}). What is the relevance of the inferred Markovian dynamics to the study of diffusive random walks in general?  We provide a speculation. The two mathematical frameworks to describe random walks,  the {\it local} Fokker-Planck  formulation and the {\it non-local} path-integral formulation are often equivalent. For example, the local assertion that all nearest neighbor jumps on an infinite regular lattice are equiprobable is equivalent to the non-local assertion that all paths of equal duration are equiprobable. But, confinement and lattice irregularities lead to prominent localization away from the boundary; a striking difference between the two approaches~\citep{burda2009localization}. This localization is usually explained by invoking fictitious entropic forces in the Fokker-Planck approach. We believe that path based approaches may  be better descriptors of stochastic dynamics especially for discrete and finite systems such as spin systems and chemical reaction networks. We leave this for future theoretical and experimental studies.

{\bf Acknowledgments:} We thank Dr. Sumedh Risbud for valuable discussions.

%\bibliography{superstat}
%\bibliographystyle{h-physrev}

\clearpage

\section{Derivation of the Markov chain}

For notational simplicity, we consider the Caliber only with one constraint $r_{ab}$. Generalization to multiple constraints is straightforward
\begin{eqnarray}
\mathcal C &=& -\sum_{a,b} p_a k_{ab} \log k_{ab} + \sum_a m_a \left (\sum_b p_a k_{ab} - p_a \right ) \nonumber \\ &+&  \sum_b n_b \left (\sum_a p_a k_{ab} - p_b \right ) + \delta \left ( \sum p_a k_{ab} - 1 \right ) \nonumber \\ &-& \gamma \left (\sum_{a,b} p_a k_{ab} r_{ab}- \langle r \rangle \right ) . \nonumber \\
\end{eqnarray}
As above, all summations involving two indices are restricted to edges of the graph.

Differentiating the Caliber with respect to $k_{ab}$, we have
\begin{eqnarray}
p_a (\log k_{ab} + 1 ) &=& p_a \left ( m_a + n_b + \delta -  \gamma r_{ab} \right ) \nonumber  \\
\Rightarrow k_{ab} &=& e^{m_a + n_b + \delta - 1 -  \gamma r_{ab} } \label{eq:kab0}
\end{eqnarray}

Differentiating the Caliber with respect to $p_a$, we have
\begin{eqnarray}
0&=&-\sum_b k_{ab} \log k_{ab} + m_a \sum_b k_{ab} - m_a+ \sum_b n_b k_{ab} - n_a \nonumber\\ &+& \delta \sum_b k_{ab} - \gamma \sum_{b}  k_{ab} r_{ab}
\end{eqnarray}

Substituting $k_{ab}$ from Eq.~\ref{eq:kab0}, we get
\begin{eqnarray}
m_a + n_a = 1
\end{eqnarray}Substituting in Eq.~\ref{eq:kab0}, we get
\begin{eqnarray}
k_{ab} = \frac{\phi_b}{\eta \phi_a} {\bf W}_{ab}
\end{eqnarray}
Here, ${\bf W}_{ab} = e^{-\gamma r_{ab}}$ when $(a,b) \in E$ and zero otherwise, $\phi_a = e^{-m_a}$, and $\eta = e^{-\delta}$. Imposing $\sum_b k_{ab} = 1$, we have
\begin{eqnarray}
\sum_b {\bf W}_{ab} \phi_b = \eta \phi_a
\end{eqnarray}
Given that ${\bf W}$ is irreducible and non-negative, it has a Perron-Frobenius eigenvalue that is positive and such that the corresponding eigenvector has positive elements. Given that the solution to the Caliber maximization problem is unique, if we choose $\bar \phi$ to be the Perron-Frobenius vector, we obtain the transition matrix elements $k_{ab}$ as
\begin{eqnarray}
k_{ab} = \frac{\phi_b}{\eta \phi_a}{\bf W}_{ab} 
\end{eqnarray}
when $(a,b) \in E$ and zero otherwise.

\section{Imposing detailed balance}

As above, we consider the Caliber
\begin{eqnarray}
\mathcal C &=& -\sum_{a,b} p_a k_{ab} \log k_{ab} + \sum_a m_a \left (\sum_b p_a k_{ab} - p_a \right ) \nonumber \\ &+&  \sum_b n_b \left (\sum_a p_a k_{ab} - p_b \right ) + \delta \left ( \sum p_a k_{ab} - 1 \right ) \nonumber \\ &+& \sum_{a,b} \epsilon_{ab} \left (p_a k_{ab} - p_b k_{ba} \right ) - \gamma \left ( \sum_{ab}   p_a  k_{ab} r_{ab}- \langle r \rangle \right ) . \nonumber \\
\end{eqnarray}
We have introduced Lagrange multipliers $\epsilon_{ab}$ to enforce detailed balance. As above, all summations involving two indices are restricted to edges of the graph.

Differentiating the Caliber with respect to $k_{ab}$, we have
\begin{eqnarray}
p_a (\log k_{ab} + 1 ) &=& p_a m_a + p_a n_b + p_a \delta + p_a (\epsilon_{ab} -\epsilon_{ba}) \nonumber \\ &-&  p_a 
\gamma r_{ab} \\
\Rightarrow k_{ab} &=& e^{\left ( m_a + n_b + \delta - 1 -\gamma r_{ab} + \epsilon_{ab} -\epsilon_{ba} \right )} \label{eq:kab1}
\end{eqnarray}

Differentiating the Caliber with respect to $p_a$, we have
\begin{eqnarray}
0&=&-\sum_b k_{ab} \log k_{ab} + m_a \sum_b k_{ab} - m_a+ \sum_b n_b k_{ab} - n_a \nonumber\\ &+& \delta \sum_b k_{ab}
 + \sum_b  k_{ab} (\epsilon_{ab} -\epsilon_{ba})  - \gamma \sum_b r_{ab} k_{ab}
\end{eqnarray}

Substituting $k_{ab}$ from Eq.~\ref{eq:kab1}, we get
\begin{eqnarray}
m_a + n_a = 1
\end{eqnarray}
Substituting in Eq.~\ref{eq:kab1}, we get
\begin{eqnarray}
k_{ab} = \frac{\alpha_b}{\eta \alpha_a} e^{-\gamma r_{ab}} \kappa_{ab}.
\end{eqnarray}
Here, $\alpha_a = e^{-m_a}$, $\eta = e^{-\delta}$, and $\kappa_{ab} = e^{\epsilon_{ab} - \epsilon_{ba}}$. Notice that $\kappa_{ab} \kappa_{ba} = 1$.

To determine $\kappa_{ab}$, we impose detailed balance,
\begin{eqnarray}
\frac{k_{ab}}{k_{ba}} &=& \frac{p_b}{p_a} = \frac{\alpha_b^2}{\alpha_a^2}e^{-\gamma r_{ab} + \gamma r_{ba}} \kappa_{ab}^2 \\
\Rightarrow \kappa_{ab} &=& \sqrt{\frac{p_b}{p_a}} \frac{\alpha_a}{\alpha_b} e^{\frac{1}{2}\gamma \left ( r_{ab} - r_{ba}\right )}
\end{eqnarray}

Thus, the transition probabilities are
\begin{eqnarray}
k_{ab} &=& \sqrt{\frac{p_b}{p_a}} \frac{\alpha_a}{\alpha_b} e^{\frac{\gamma}{2} \left ( r_{ab} - r_{ba}\right )} \frac{\alpha_b}{\eta \alpha_a} e^{-\gamma r_{ab}} \\
&=& \frac{1}{\eta}\sqrt{\frac{p_b}{p_a}} e^{-\frac{1}{2}\gamma\left ( r_{ab} + r_{ba} \right )}
\end{eqnarray}

Let $\phi_a = \sqrt{p_a}$ and ${\bf W}_{ab} = e^{-\frac{1}{2}\gamma\left ( r_{ab} + r_{ba} \right )}$ when $(a,b) \in E$ and zero otherwise. Using $\sum_b k_{ab} = 1$, we have
\begin{eqnarray}
\sum_b {\bf W}_{ab} \phi_b = \eta \phi_a
\end{eqnarray}
Thus, $\bar \phi$, the vector of square roots of probabilities is the eigenvector of {\bf W} with eigenvalue $\eta$. Thus, imposing detailed balance is equivalent to constraining a symmetrized form of the constraints.

\end{document}